
\def\ccc{\chi}
\def\ddd{\delta} 
 
\def\ve{\varepsilon}

\def\lll{\lambda} 
\def\mmm{\mu} 
\def\nnn{\nu}
\def\ooo{\omega} 
\def\ppp{\pi} 

\def\qqq{\theta}

\def\ttt{\tau}

\def\yyy{\psi} 
\def\ybf{{\yyy\!\!\!\!\yyy\!\!\!\!\yyy\!\!\!\!\yyy}}

\def\F{\Phi} 
\def\G{\Gamma}

\def\P{\Pi}
\def\Q{\Theta}

\def\Y{\Psi}

\def\({\left(}
\def\){\right)}
\def\[{\left[}
\def\]{\right]}

\def\2#1#2{\la#1, #2\ra} 
\def\8{\infty}
\def\={\equiv}

\def\b#1{\bar#1}
\def\c#1{{\cal#1}} 
\def\cc#1{{{\bf C}^#1}}
 
\def\cx{{\bf C}} 
\def\curl{\nabla\times} 
\def\d#1#2{d^#1\v#2\,} 
\def\dbb{d\kern-1.5ex-\kern-.4ex} 
\def\db{d\kern-2ex-\kern-.4ex} 

\def\del{{\sqcup\hskip-1.6ex\sqcap\ }}
 
\def\div{\nabla\cdot } 
\def\dpt{d\t p\ }

\def \grad{\nabla}
\def\h#1{\hat#1}

\def\i1#1{\int_{-\infty}^\infty d#1\,\,} 
\def\ii{^{-1}}

\def\ir{\int_{-\infty}^\infty}

\def\la{\langle\, }

\def\M{\,|\,} 
\def\mt{\mapsto}
 
\def\notto{\ /\kern-2.2ex \to }
\def\0#1{(#1)} 
\def\ol#1{\overline{#1}}

\def\pl{\partial} 

\def\q{\quad} 
\def\qq{\qquad} 
\def\qed{\vrule height6pt width3pt depth 0pt}

\def\ra{\,\rangle} 
\def\rat#1#2{{{#1}\over{#2}}}
\def\re{{\rm Re}\,}   
\def\ref{\null}
 
\def\rr#1{{{\bf R}^#1}}
\def\sh#1{\hglue#1ex} 
\def\sp#1#2{\v #1\cdot\v #2\,}

\def\st#1{{\sl{#1}\/}} 
\def\sv#1{\vglue#1ex}
\def\tp{{(2\pi )}}  
\def\t#1{\tilde#1} 
\def\tp{{(2\ppp)}}

\def\v#1{{\bf#1}}
 
\def\vh#1{{\bf{\h#1}}}

\def\x{\null}

\def\cl{\centerline}

\def\ff{following} 
\def\fs{follows} 
\def\FT{Fourier transform} 
\def\hs{Hilbert space} 
\def\ie{i.e., }
\def\iff{if and only if}
 
\def\ip{inner product}
    
\def\n{\noindent}

\def\resp{respectively}
\def\rhs{right--hand side}  
\def\ru{resolution of unity}

\def\wrt{with respect to}

\def\e#1{\eqno(#1.\the\eq)\global\advance\eq by 1$$}
\def\en{\eqno(\the\eq)\global\advance\eq by 1$$}

\def\refs{\bigbreak{\centerline{\bf References}}
\bigskip\frenchspacing\everypar={\hangindent=2em\hangafter
=1}\parindent=0pt\parskip=\smallskipamount}



\magnification=1200   
\hsize=16 true cm
\vsize=21 true cm  
\hoffset -.2 cm 
\newcount\eq\eq=1 

\vglue 2ex
\baselineskip=12pt 

\centerline{\bf Wavelet Electrodynamics II:}

\vglue 1ex

\centerline{\bf Atomic Composition of Electromagnetic Waves}

\sv2

\cl{(Appeared in \sl Applied and Computational Harmonic Analysis \rm {\bf 1}, 246--260,
1994)}

\vglue 3ex

\centerline{\bf Gerald Kaiser} 
\centerline{Department of Mathematical Sciences} 
\centerline{University of Massachusetts at Lowell} 
\centerline{Lowell, MA 01854, USA} 
\centerline{e-mail: kaiserg@ woods.ulowell.edu}

\vglue 2ex

\centerline{November 15, 1993}

\sv4

\centerline{\bf Abstract} 
\sv1

\noindent The representation of  solutions of Maxwell's equations 
as  superpositions of scalar wave\-lets with vector coefficients
developed earlier is generalized to wavelets with polarization,
which are matrix-valued.  The construction  proceeds in four stages:
 {\bf(1)} A Hilbert space $\c H$ of solutions is considered, based
on a conformally invariant \ip. {\bf(2)} The \st{analytic-signal
transform} is used to extend solutions  from real space-time to a
complex  space-time domain $\c T$.   The evaluation maps
$\c E_z$, which send any solution $\v F=\v B+i\v E$ to the  values 
$\t\v F(z)$ of its extension  at points $z\in\c T$, are bounded linear
maps on $\c H$. Their adjoints $\v \Y_z\=\c E_z^*$ are the 
electromagnetic wavelets.   {\bf(3)} The eight real parameters
$z=x+iy\in \c T$ are given a complete physical interpretation: 
$x=(\v x, t)\in\v R^4$ is interpreted as a space-time point about
which $\v \Y_z$ is \st{focussed.} The imaginary space-time vector
$y=(\v y, s)$ is time-like, i.e.,  $|\v y|<|s|$.  The sign of $s$  is
interpreted as the \st{helicity} of the wavelet, while $|s|$ is its 
\st{scale.}   The 3-vector $\v v\=\v y/s$ is the velocity of its
center.   Thus wavelets parameterized by the set of {\sl
Euclidean\/}  points $E=\{(\v x, is)\}$ (real space and imaginary
time)  have stationary centers, and wavelets with $\v y\ne\v 0$
are Doppler-shifted versions of ones with stationary centers.  All
the wavelets can be obtained from a single ``mother'' by conformal
transformations. {\bf(4)}  A resolution of unity is established in $\c
H$,  giving a representation of solutions  as ``atomic compositions''
of wavelets  parameterized by $z\in E$.  This yields a constructive
method for generating solutions with initial data specified locally
in space and by scale.  Other representations, employing wavelets
with moving centers,  are obtained by applying conformal
transformations to the stationary representation.  This could be
useful in the analysis of electromagnetic waves reflected or
emitted by moving objects, such as radar signals.

\vfill\eject

\n{\bf 1. Introduction}

\sv1

\n In this paper we further develop the wavelet formulation of
classical electrodynamics begun in Kaiser \ref[12-14].  There, it was
shown that electromagnetic waves (solutions of Maxwell's
equations) can be expressed as linear superpositions of spherical
wavelets uniquely adapted to these equations.  However, the
wavelets  in Refs.~\ref[12-14] were scalar-valued
solutions of the wave equation rather than (vector-valued)
solutions of Maxwell's equations.  (Their coefficients were
vector-valued, thus providing for polarization.)
Although the scalar wavelets sufficed for the \st{reconstruction} of
known electromagnetic waves, they could not be used for the
\st{construction} of new ones according to given local
data, precisely because of their scalar nature, which ignored the
polarization degrees of freedom. (Their reproducing kernel was not
the projection operator to the solution space.)  The wavelets
constructed here remedy this deficiency, since they are true
electromagnetic fields, parameterized by a complete set of
physically relevant variables: Their point and time of localization,
the velocity of their center, their scale and their helicity.

This work is part of a general program
whose main objective is to \st{extend the elementary physical
fields from real to complex space-time and interpret the
imaginary parts of the space-time variables as control parameters
for the wave number and frequency contents of the fields
being analyzed.}  This approach has so far given a fully
relativistic phase-space description of a variety of free
field theories:  Klein-Gordon, Dirac, and now Maxwell. 
Furthermore,   the regularity resulting from the
analyticity of the fields in the complex space-time domain
promises to  help resolve some of the fundamental difficulties
plaguing these theories,  related to their reliance on the concept of
precise geometrical points with no substance or structure.  For the
extended fields, \st{points in complex space-time have a natural
interpretation as moving extended objects in real space-time
which, in turn, act as elementary building blocks or ``atoms'' for
the fields.}  In the case of field theories with positive mass
(Klein-Gordon, Dirac), these atoms  are relativistic coherent states,
\ie Gabor-like wavelets  whose windows  undergo scaling
(Lorentz contractions) under Lorentz transformations. In the case of
massless field theories such as electrodynamics, the atoms are
space-time-scale wavelets transforming covariantly under  the
conformal group.  In all cases, the atoms are uniquely determined
by the field theory through covariance and analyticity.

 The main ideas of the above program were reported in \ref[16],
where some of the results of the author's thesis (\st{Phase-Space
Approach to Relativistic Quantum Mechanics,} Univ.~of Toronto,
1977) were summarized.  A key tool for extending general fields to
complex space-time, the \st{analytic-signal transform,} was
developed in \ref[17] and further investigated in \ref[11] .   Some
speculation on the application of these methods to electrodynamics
was advanced in Kaiser and Streater \ref[15]  for the much simpler
case of a two-dimensional space-time.

In Section \x2,  Maxwell's equations are solved  by Fourier
methods from a viewpoint in which  the concepts of helicity and
polarization become very clear.   A Hilbert space $\c H$ of solutions
is constructed which was proved by L.~Gross \ref[8] to carry a
unitary representation of the full invariance group of the equations,
namely the conformal group of space-time.   In Section \x3 we
review the analytic-signal transform, which extends any function
$f\0x$ from $\rr n$ to $\cc n$.  In general, the extended function
$\t f\0z$ is not analytic (there may not \st{exist} any analytic
extension).  But when the \FT{}  of $f$ is supported on a double
convex cone (as it is,  for example, when $f$ represents a  free
relativistic field such as an electromagnetic wave),  then $\t f\0z$
is  analytic in a certain tube domain $\c T$ in $\cc n$.  In Section
\x4 we show that the analytic-signal transform, when applied to
electromagnetic fields, uniquely determines a set of
electromagnetic wavelets. A \ru{} is derived which allows
 solutions to be expressed as superpositions of wavelets. 
In Section \x5 we compute  the  reproducing kernel
defined by the  wavelets, which in turn gives  the wavelets
explicitly.  In Section \x6 we show how arbitrary solutions in $\c
H$ can be constructed from wavelets, with initial data specified
locally and by scale.  In Section \x7,  the  wavelets are given a
complete physical and geometric interpretation.    In Section \x8
we describe some generalizations, and  a possible application.

\sv4

 \n{\bf  2.  The Fourier Representation of Free
Electromagnetic Fields}

\sv1

\n  An electromagnetic wave in free space (without sources or
boundaries) is described by a pair of vector  fields depending on
the space-time variables $x=(\v x, x_0)$ (where $\v x$ is the
position and $x_0$ is the time), namely the electric field $\v E\0x$
and the magnetic field $\v B\0x$.  These are subject to Maxwell's
equations,

$$
\eqalign{ \curl\v E+\pl_0\v B=\v 0,\qq & \div\v E=0,  \cr
\curl\v B-\pl_0\v E=\v 0, \qq &\div \v B=0, 
\cr}\en 
where $\nabla$ is the gradient \wrt{} the  space variables and
$\pl_0$ is the time derivative.  We have set the speed of light
$c=1$ for convenience.  For the present dicussion, $\v E$ and $\v
B$ may be assumed to be tempered distributions, so that \x(2.1)
holds weakly.  Later, the fields will be required to belong to a 
certain Hilbert space.  Note that the equations are symmetric
under the linear mapping defined by $J: \v E\mt \v B,\,\, \v B\mt
-\v E$, and that $J^2$ is minus the identity map. 
Such a mapping on a vector space is called a \st{complex
structure,}  by analogy with multiplication by $i$ in the complex
plane.   The combinations $\v B\pm i\,\v E$ diagonalize $J$, since 
$J(\v B\pm i\,\v E)=\pm i\,(\v B\pm i\,\v E)$.  They each map
Maxwell's equations to a form in which the concepts of  helicity and
polarization become very simple.  It will suffice to consider only
$\v F\=\v B+i\,\v E$, since the other combination is equivalent. 
Eqs.~\x(2.1) then become

$$
\pl_0\v F=i\curl\v F,\qq \div\v F=0. 
\en 
Note that the first of these equations is an evolution equation
(initial-value problem), while the second is a constraint on the
initial values.  Taking the divergence of the first equation shows
that the constraint is conserved by time evolution.  Note also that
it is the factor $i$ in \x(2.2) (\ie the complex structure!)
which couples the dynamics of the electric field $\v E$ to those of
the magnetic field $\v B$.   Eq.~\x(2.2) implies

$$
-\pl_0^2\v F =\curl(\curl\v F)
=\grad(\div\v F)-\nabla^2\v F=-\nabla^2\v F
\en 
in Cartesian coordinates, hence the components of $\v F$ become
decoupled  and each satisfies the wave equation 

$$ 
\del\v F\=(-\pl_0^2+\nabla^2)\v F=\v 0. 
\en 
Since $\v F$ is a tempered distribution, it has a Fourier expansion

$$
\v F\0x=(2\ppp)^{-4}\int_\rr4 d^4p \,e^{ip\cdot x}\,\h\v F\0p,
\en
where  $p=(\v p, p_0)\in\rr4$ with  $\v p\in\rr3$ as the spatial
wave vector and $p_0$ as the frequency.  We use the
Lorentz-invariant \ip{}  $p\cdot x\=p_0 x_0-\v p\cdot\v x$.  The
wave equation \x(2.4) implies that  $p^2\h\v F\0p=0$, where
$p^2\=p\cdot p=p_0^2-|\v p|^2$.  If the distribution 
$\h\v F$ has no essential support at the origin $p=0$ (\ie no
term proportional to $\ddd\0p$),   it must be supported  on the
nipped light cone

$$ 
C=\{(\v p, p_0)\in\rr4\backslash\{0\}:  p_0^2=|\v p|^2\}
=C_+\cup C_-,  
\en  
where $\pm p_0=|\v p|>0$ in $C_\pm$.  Hence $\vh F$ has the
form $\vh F(p)=2\ppp\ddd(p^2)\,\v f(p)$,  where $\v f\0p$ is a
(vector-valued) distribution on $C$ or, equivalently,  the \st{pair}
of  distributions on $\rr3$  given by $\v f_\pm(\v p)
\=\v f(\v p, \pm |\v p|)$.  For the moment,  we assume that $\v
f_\pm$ are (vector-valued) Schwartz test functions. 
Later the class of $\v f$'s will be enlarged by introducing an \ip{}
and completing it to a Hilbert space, subject to a restriction related
to our having ``nipped'' the light cone, which amounts, roughly,  to 
$\v f\00=\v 0$.  Letting $\ooo\=|\v p|$,  we have

$$
\ddd(p^2)=\ddd((p_0-\ooo)(p_0+\ooo))
=\rat{\ddd(p_0-\ooo)+\ddd(p_0+\ooo)}{2|p_0|}.
\en
Hence \x(2.5) becomes

$$\eqalign{
\v F\0x
&=\tp^{-3}\int_\rr3 \rat{\d3p}{2|p_0|}\,e^{-i\v p\cdot\v x}\,
\[e^{i\ooo t}\,\v f_+(\v p)+e^{-i\ooo t}\,\v f_-(\v p) \]  \cr
&=\int_C\dpt e^{ip\cdot x} \,\v f\0p, 
\cr}\en 
where  $\dpt\=\tp^{-3}d^3\v p/2\,|p_0|$ is a Lorentz-invariant
measure on $C$.  In order for \x(2.8) to give a solution of \x(2.2),
$\v f\0p$ must further satisfy the algebraic conditions

$$ 
i\,p_0\,\v f\0p=\v p\times\v f\0p,\qq \v p\cdot\v f\0p=0 
\en 
 for
all $p\in C$,  and the first of these equations suffices since it
implies the second.  Let $\v v\0p=\v p/p_0$, so that $p\in C$ \iff{}
$|\v v|=1$.   Define the operator $\v\G\=\v\G\0p$
on arbitrary functions $\v g: C\to\cc3$ by

$$
 (\v\G\v g)\0p\=-i\,\v v\0p\times \v g\0p,\qq p\in C. 
\en 
$\v\G\0p$
is represented by the Hermitian matrix

$$ 
\v\G\0p=i\[\matrix{0&v_3&-v_2\cr
                             -v_3&0&v_1\cr
                             v_2&-v_1&0\cr}\]
=\rat1{p_0}\[\matrix{0&p_3&-p_2\cr
                             -p_3&0&p_1\cr
                             p_2&-p_1&0\cr}\], 
\en 
 with matrix elements
$\G_{mn}\0p=ip_0\ii\sum_{k=1}^3\ve_{mnk}\,p_k\,$, where
$\ve_{mnk}$ is the totally antisymmetric tensor with
$\ve_{123}=1$. In terms of $\v\G\0p$, \x(2.9) becomes 

$$ 
\v\G\v f\0p=\v f\0p.
\en 
 Now for any $\v g: C\to\cc3$,

$$ 
\v\G^2\v g
=-\v v\times(\v v\times \v g) 
=\v g-(\sp vg)\v v, 
\en 
 so $\v\G\0p^2$ is the orthogonal projection to the
subspace of $\cc3$ orthogonal to $\v v\0p$, and it \fs{} that

$$ 
\v\G^3\v g=\v\G\v g.
\en  
The eigenvalues of $\v\G\0p$, for each $p\in C$, are therefore $1, 0$
and $-1$, and \x(2.12) states that $\v f\0p$ is an eigenvector with
eigenvalue 1.   Since $\ol{\v\G\0p}=-\v\G\0p$, \x(2.12) implies
that $\v\G\b f=-\b f$.  A similar operator was defined and
studied in much more detail by Moses \ref[19]  in connection with
fluid mechanics as well as electrodynamics.

Consider a single component of \x(2.8), \ie the plane-wave
solution

$$
\v F_p\0x\=e^{ip\cdot x}\,\v f\0p=\v B_p\0x+i\,\v E_p\0x,
\en
with arbitrary but fixed $p\in C$ and $\v f\0p\ne\v 0$.
The electric and magnetic fields are obtained by taking the real and
imaginary parts.  Now $\v\G\0p\,\v f\0p=\v f\0p$ and
$\v\G\0p\,\ol{\v f\0p}=-\ol{\v f\0p}$  imply

$$
\v\G\0p\v F_p\0x=\v F_p\0x,\qq  
\v\G\0p\,\ol{\v F_p\0x}=-\ol{\v F_p\0x}.
\en
Since $\v\G\0p^*=\v\G\0p$, these eigenvectors of $\v\G\0p$ 
 with eigenvalues 1  and $-1$ must be orthogonal:
$\ol{\v F_p(x})^*\v F_p\0x=\v F_p\0x\cdot\v F_p\0x=0$,
where the asterisk denotes the Hermitian transpose.  Taking
real and imaginary parts, we get

 $$
|\v B_p\0x|^2=|\v E_p\0x|^2,\qq
\v B_p\0x\cdot\v E_p\0x=0.
\en
The first equation shows that neither $\v B_p\0x$ nor $\v E_p\0x$
can vanish at any $x$ (since $\v f\0p\ne \v 0$).  Furthermore, 
\x(2.16)  implies that  $\v p\times \v E_p\0x=p_0\v B_p\0x$.
Thus, for any $x$,  $\{\v p, \v E_p\0x, \v B_p\0x\}$ is a
\st{right-handed orthogonal basis} if $p_0>0$ (\ie $p\in C_+$) and 
a \st{left-handed orthogonal basis}  if $p_0<0$  ($p\in C_-$). 
Taking the real and imaginary parts of  \x(2.15) and using 
$\v f\0p=\v B_p\00+i\,\v E_p\00$, we have

$$\eqalign{
\v B_p\0x
&=\cos (p\cdot x)\v B_p\00-\sin (p\cdot x)\v E_p\00,\cr 
\v E_p\0x
&=\cos (p\cdot x)\v E_p\00+\sin (p\cdot x)\v B_p\00.
 \cr}\en
An observer at any fixed location $\v x\in\rr3$ sees
these fields rotating, as a function of time, in the plane orthogonal
to $\v p$.  If $p\in C_+$, the rotation is  that of a right-handed
corkscrew, or helix,  moving in the direction of $\v p$,  whereas if
$p\in C_-$, it is that of a left-handed corkscrew.  Hence $\v
F_p\0x$ is said to have \st{positive helicity} if $p\in C_+$ and  
\st{negative helicity} if $p\in C_-$.

A general solution of the form \x(2.8)  has positive helicity if $\v
f\0p$ is supported in $C_+$ and negative helicity if $\v f\0p$ is
supported in $C_-$.  Other states of polarization, such as linear or
elliptic, are obtained by mixing positive and negative helicities.  The
significance of the complex combination $\v F\0x=\v B\0x+i\,\v
E\0x$ therefore seems to be that \st{in Fourier space,  the sign of
the frequency $p_0$  gives the helicity of the solution!} (Usually in
signal analysis, the sign of the frequency is not given any physical
interpretation, and negative frequency is regarded as a convenient
mathematical artifact.) In other words,  the combination $\v B+i\,\v
E$ ``polarizes'' the helicity, with positive and negative helicity
states being represented in $C_+$ and $C_-\,$, \resp. Had we used
the opposite combination $\v B-i\,\v E$,  $C_+$ and $C_-$ would
have parameterized the plane-wave solutions with opposite
helicities.  Nothing new seems to be gained by considering this
alternative.  (In fact, Maxwell's equations are invariant under the
continuous group of \st{duality rotations,} of which the complex
structure $J$ mapping $\v E$ to $\v B$ and $\v B$ to $-\v E$ is a
special case.  In the complexified solution space, the combinations
$\v B\pm i\,\v E$ form invariant subspaces with respect to the
duality rotations.  That gives the choice of $\v B+i\v E$ an
interpretation in terms of  group representation theory.)

In order to eliminate the constraint, we now proceed as
\fs: Let 

$$ 
\v\P\0p\=\rat12\[\v\G\0p+\v\G^2\0p\].
\en 
Explicitly,

$$ 
\v\P\0p=\rat1{2p_0^2}\,\[\matrix{
p_0^2-p_1^2&-p_1p_2+ip_0p_3&-p_1p_3-ip_0p_2\cr
-p_1p_2-ip_0p_3&p_0^2-p_2^2&-p_2p_3+ip_0p_1\cr
-p_1p_3+ip_0p_2&-p_2p_3-ip_0p_1&p_0^2-p_3^2\cr}\].
\en 
The established properties $\v\G^*=\v\G=\v\G^3$ imply that 
$\v\P^*=\v\P=\v\P^2$ and $\v\G\v\P=\v\P$, which proves that $\v\P\0p$ is the
orthogonal projection to eigenvectors of $\v\G\0p$ with eigenvalue
1. Thus, to satisfy the constraint \x(2.12),  we need only replace the
constrained function $\v f\0p$ in \x(2.8) by $\v\P\0p\v f\0p$,
where now  $\v f\0p$ is unconstrained: 

$$ 
\v F\0x
=\int_C\dpt e^{ip\cdot x} \,\v\P\0p\,\v f\0p\,.
\en 
Consequently,  the mapping $\v f\mt\v F$ is not one-to-one since
$\v\P$ is a projection operator. In fact,  $\v f$ is closely related to the
\st{potentials} for $\v F$, which consist of a real 3-vector
potential  $\v A\0x$ and a real scalar potential $A_0\0x$  such
that  

$$
\v B=\curl \v A, \qq\v E=-\pl_0\v A-\grad A_0.
\en
The combination $(\v A\0x, A_0\0x)$ is called a  ``4-vector
potential'' for the field.  We can  assume without loss of
generality that the potential satisfies the Lorentz condition\break
  $\div \v A+\pl_0A_0=0$ (Jackson [10]). Since $\v A$ and $A_0$
also satisfy the wave equation \x(2.4), they have  Fourier
representations similar to \x(2.8):

$$
\v A\0x=\int_C\dpt \,e^{ip\cdot x}\,\v a\0p,\q
A_0\0x=\int_C\dpt \,e^{ip\cdot x}\,a_0\0p.
\en
The Lorentz condition means that $p_0 a_0\0p=\v p\cdot \v
a\0p$, or $a_0\0p=\v v\cdot \v a\0p$, so $a_0$ is determined by
$\v a$.  Eqs.~\x(2.22) will be satisfied provided that the Fourier
representatives $\v e\0p,\,  \v b\0p$ of $\v E, \v B$ satisfy

$$
 \v b=-i\,\v p\times\v a=p_0\,\v\G\,\v a,\qq
\v e=-ip_0\v a+i\,\v p a_0 =-ip_0\,\v\G^2\,\v a.
\en
Hence  $\v F=\v B+i\,\v E$ is represented in Fourier space by

$$
\v b\0p+i\,\v e\0p=p_0\[\v\G\0p+\v\G\0p^2\]\,\v a\0p
=2p_0\v\P\0p\,\v a\0p. 
\en
 This shows that we can interpret the unconstrained function $\v
f\0p$ in \x(2.21) as being directly related to the 3-vector potential
by

$$
\v f\0p=2p_0\,\v a\0p,
\en
modulo terms annihilated by $\v\P\0p$, which correspond to
eigenvalues $-1$ and 0 of $\v\G\0p$.  Seen in this light, the
non-uniqueness of $\v f$ in \x(2.21) is  an expression of \st{gauge
freedom} in the $\v B+i\,\v E$ representation, as seen from Fourier
space.  In the space-time domain, $(\v B,\,\v E)$  are the
components of a 2-form $F$ in $\rr4$ and $(\v A, A_0)$ are the
components of a 1-form $A$.  Then Eqs.~\x(2.1)  become 
$d\,F=0$ and $\ddd F=0$ (where $\ddd$ is the divergence \wrt{}
the Lorentzian \ip),  Eqs.~\x(2.22) become unified as $F=dA$,  the
Lorentz condition reads $\ddd A=0$  and the gauge freedom
corresponds to the invariance of $F$ under $A\to A+d\ccc$, where
$\ccc\0x$ is a scalar solution of the wave equation.

Maxwell's equations are invariant under a large group of
space-time transformations.  Such transformations produce new
solutions from known ones by acting on the underlying space-time
variables (possibly with a multiplier to rotate or scale the vector
fields).  Some trivial examples are space and time translations:
Obviously, a translated version of a solution is again a solution,
since the equations have constant coefficients.  Similarly, a rotated
version of a solution is a solution.  A less obvious example is
Lorentz transformations, which are interpreted as transforming to a
uniformly \st{moving} reference frame in space-time.  (In fact, it
was in the study of the Lorentz  invariance of Maxwell's equations
that the Special Theory of Relativity originated;  see Einstein et al.
\ref[6].)  The scale transformations $x\to ax,\ a\ne 0$, also map
solutions to solutions, since Maxwell's equations are homogeneous
in the space-time variables. Finally,  the equations are  invariant
under ``special conformal transformations'' (Bateman \ref[2],
Cunnigham \ref[4]),  which can be interpreted as transforming to a
uniformly \st{accelerating} reference frame (Page \ref[20]; Hill
\ref[9]).  Altogether, these transformations form a 15-dimenional
Lie group called the \st{conformal group,} which  is locally
isomorphic to $SU(2,2)$ and is here denoted by $\c C$.  Whereas
wavelets in one dimension are related to one another  by
translations and scalings, electromagnetic wavelets will be seen to
be related by conformal transformations, which include
translations and scalings.  (A study of the action of $SU(2,2)$ on 
solutions of Maxwell's equations has been made by R\" uhl [21].)

To construct the machinery of wavelet analysis, we  introduce a
Hilbert space structure on the solutions.  It is important to choose
the \ip{} to be invariant under the largest possible group of
symmetries, since this allows the largest set of solutions in $\c H$
to be generated by unitary transformations from any one known
solution.  (In quantum mechanics,  invariance of the \ip{}  is also
an expression of the fundamental invariance of the laws of nature
\wrt{} the symmetries in  question.)  Let $\v f\0p$ satisfy
\x(2.12), and let $\v a\0p$ be a vector potential for $\v f$
satisfying the Lorentz condition, so that the scalar potential is
determined by $a_0\0p=\v v\cdot \v a\0p$.  By \x(2.25),

$$\eqalign{
|\v f\0p|^2
&=4p_0^2\,|\v\P\0p\,\v a\0p|^2
=4p_0^2\,\,\ol{\v a\0p}\cdot  \v\P\0p\,\v a\0p\cr 
&=2p_0^2\,\,\ol{\v a\0p}\cdot \v\G\0p\,\v a\0p 
+ 2p_0^2\,\,\ol{\v a\0p}\cdot\v\G\0p^2\,\v a\0p. 
\cr}\en
The first term is 

$$
-2ip_0^2\,\,\ol{\v a\0p}\cdot(\v v\times\v a\0p)
=2ip_0^2\,\v v\cdot(\ol{\v a\0p}\times\,\v a\0p),
\en
which cancels its counterpart with $p\to -p$ on account of the
reality condition $\ol{\v a(-p)}=\v a\0p$.  Thus

$$\eqalign{
\int_C\rat\dpt{p_0^2}\,|\v f\0p|^2
&=2\int_C\dpt\,\,\,\ol{\v a\0p}
\cdot \bigl[\v a\0p-\v v(\v v\cdot\v a\0p)\bigr] \cr
&=2\int_C\dpt\,\Bigl[|\v a\0p|^2-|a_0\0p|^2\Bigr].
\cr}\en
The integrand in the last expression is the negative of the
Lorentz-square of the 4-potential $(\v a\0p, a_0\0p)$. 
Consequently, the  integral can be shown to be invariant  under
Lorentz transformations. (Note that  $|\v a|^2-|a_0|^2 \ge 0$,
vanishing only when $\v a\0p$ is a  multiple of $\v p$,  in which
case $\v f=\v 0$.  This corresponds to ``longitudinal polarization.'')
Hence \x(2.29) defines a norm on solutions which is invariant
under Lorentz transformations as well as space-time translations. 
In fact,  the norm \x(2.29) is uniquely determined, up to a constant
factor, by the requirement that it be so invariant.
Moreover, Gross [8] has shown it to be invariant under the full
conformal group $\c C$.  Again we eliminate the constraint by
replacing $\v f\0p$ with $\v\P\0p\,\v f\0p$. Thus, let $\c H$ be
the set of all solutions $\v F\0x$ defined by \x(2.21) with $\v f:
C\to \cc3$  square-integrable in the sense that

$$ 
\|\v F\|^2 =\int_C\rat\dpt{p_0^2}\,|\v\P\0p \,\v f\0p|^2
=(2\ppp)^{-3}\int_C\rat {\d3p}{2|\v p|^3}\,|\v\P\0p\,\v f\0p|^2 <\8.
\en 
$\c H$ is a Hilbert space under  the \ip{} obtained by
applying the polarization identity to \x(2.30) and using
$(\v \P\v f)^*\v \P\v g=\v f^*\v \P^*\v \P\v g=\v f^*\v \P\v g$:

$$
\la \v F, \v G\ra=\int_C\rat\dpt{p_0^2}\,\v f\0p^*\v\P\0p\,\v g\0p.
\en
$\c H$ will be our main arena for developing the wavelet
analysis  and synthesis of solutions.  Note that when \x(2.12) holds
and $\v f_\pm(\v p)\=\v f(\v p, \pm|\v p|)$ are Schwartz
test function as we  assumed earlier, then 

$$
\v f\00=\v f_\pm(\v 0)=\v 0
\en
must hold in order that \x(2.30) be satisfied.  However, now that
we have our Hilbert structure,  we complete to the larger class of
all (generalized) functions $\v f$ satisfying \x(2.30).

 To show the invariance of \x(2.30) under conformal
transformations, Gross derived an equivalent norm expressed 
directly in terms of the values of the fields in space at any
particular time $x_0=t$:

$$ 
\|\v F\|^2_{\rm Gross}
\= \rat1{\ppp^2}\int_\rr6\rat{\d3x\d3y }
{|\v x-\v y|^2} \  \v F(\v x, t)^*\v F(\v y, t) . 
\en 
The \rhs{} is independent of $t$ due to the invariance of Maxwell's
equations under time translations (which is, in turn,  related to
the conservation of energy).  A disadvantage of the expression
\x(2.33) is that it is \st{non-local,} since it uses the values of the
field simultaneously at the space points $\v x$ and $\v y$.  In fact,
it is known that \st{no local expression for the \ip{} can exist in
terms of the field values in (real) space-time} $\rr4$ (Bargmann
and Wigner \ref[1]).  In  Section \x4, we derive an alternate
expression for the \ip{} directly in terms of the values of the
electromagnetic fields,   extended analytically to \st{complex}
space-time.  This expression is ``local''  in the space-scale domain
(rather than in space alone).  But first we must introduce the tool
which  implements the extension to complex space-time.

\sv4

\newcount\eq\eq=1

\n{\bf 3. The Analytic-Signal Transform}

\sv1

\n Given a vector function $\v F: \rr n\to\cc m$, we define its
\st{analytic-signal transform}  as the function $\t\v F: \cc
n\to\cc m$ given by the \ff{} line integral in $\rr n$:

$$ 
\t\v F(x+iy)
=\rat1{\ppp i}\ir\rat{d\ttt}{\ttt-i}\, \v F(x+\ttt y).
\e3
 This transform was introduced and studied in Kaiser 
\ref[11, 17],  where it was shown to be related to the \FT{}
$\h\v F(p)$ of $\v F(x)$ by

$$ 
\t\v F(x+iy)
=\tp^{-n}\int_{\rr n}d^n p\, 
2\qqq( p\cdot y)\,e^{i p\cdot( x+i y)}\,\h\v F( p)\,.
\e3 
Here $\qqq$ is the unit step function, defined by $\qqq(u)=0$ if
$u<0$, $\qqq\00=\rat12$, $\qqq\0u=1$ if $u>0$.  For \x(3.2) to
make sense, it suffices that $\h\v F\0p$  be absolutely integrable,
since  $|\qqq( p\cdot y)\,e^{i p\cdot( x+i y)}|\le 1$.  For
concreteness, we assume for the time being that $\h\v F\in
L^1(\rr4)$ and use \x(3.2) to \st{define} $\t\v F$, viewing \x(3.1)
as ``motivation.''  (A study of \x(3.1) in the context of distribution
theory is currently being  undertaken by T.~Takiguchi [24].)
Note that setting $y= 0$ on the right  gives the
inverse \FT{} of $\h\v F( p)$, so that \st{formally} we have $\t\v
F( x)=\v F( x)$ and $\t\v F$ is an \st{extension} of $\v F$ from
$\rr n$ to $\cc n$.  (This is made more precise below, in \x(3.7).) 
Of course, this extension is usually not analytic, since in general
there \st{exists} no  analytic extension.  However, when $\rr n$ is
space-time and $\v F$ represents a free physical field such as an
electromagnetic field ($m=3$), a Klein-Gordon field ($m=1$) or a
Dirac field ($m=4$), then $\h\v F (p)\=\h\v F(\v p, p_0)$ vanishes
outside the solid light cone 

$$
 V\=\{(\v p, p_0)\in\rr4: p^2
\= p_0^2-|\v p|^2\ge 0,\  p_0\ne 0\}
=V_+\cup V_-,  
\e3 
 where $\pm p_0\ge |\v p|>0$ in $V_\pm$. (In
the electromagnetic case, for example, $\h\v F$ is  supported on the
\st{boundary} $C$ of $V$ as a consequence of the wave equation
\x(2.4).)  Hence the integral in \x(3.2) extends only over $V$.  
Formally, the obstacle to the analyticity of $\t\v F\0z$ in \x(3.2) is
the factor $\qqq(p\cdot y)$ (which is necessary, generally, to
ensure that the integral converges in the region of Fourier space
where  $e^{-p\cdot y}>1$).   However, when $\h\v F$ is supported
in $V$,  that obstacle can be removed as follows:  Suppose $y$ is
such that  $p\cdot y>0$ for all $p\in V_+$ and $p\cdot y>0$ for all 
$p\in V_-$.  (This means that the hyperplane Lorentz-orthogonal to
$y$ \st{separates} $V_+$ and $V_-$!)
 Then $\qqq(p\cdot y)=1$ for all $p\in V_+$ and 
$\qqq(p\cdot y)=0$ for all $p\in V_-$, hence the integral now
extends only over $V_+$, and the obstracting factor is 
identically =1 in that cone.  Furthermore, the extra factor
$e^{-p\cdot y}$ coming from the analytic continuation of the
Fourier kernel provides exponential damping, which leads to the
analyticity of $\t\v F$ at  $(x+iy)$, for all $x$.  Similarly, if $y$ is
such that $p\cdot y>0$ for all $p\in V_-$ and $p\cdot y<0$ for all
$p\in V_+$, then $\t\v F$ is again analytic at $x+iy$, for all $x$. 
The above sets of imaginary space-time points $y$ are, by
definition, the \st{dual cones} $V'_\pm$ of $V_\pm$ (Stein and
Weiss [23]),

$$
 V'_\pm\=\{y=(\v y, y_0)\in\rr 4: p\cdot y
\=p_0y_0-\v p\cdot\v y>0 \ \hbox{for all}\ p\in V_\pm\}. 
\e3 
 $V'_+$ and $V'_-$ are the  \st{open future light cone} and the
\st{open past light cone} in space-time (as opposed to Fourier
space, where $V_\pm$ live).  The union  $V'\=V'_+\cup V'_-$ will
be called the dual cone of $V=V_+\cup V_-$.  Explicitly,

$$
V'_\pm=\{(\v y, y_0): \pm y_0>|\v y|\}, \q
V'=\{y\in\rr4: y^2\=y_0^2-|\v y|^2>0\}.
\e3
The argument used above to motivate the definition of
$V'_\pm$ can be made precise, leading to the conclusion that $\t\v
F\0z$ is indeed analytic in

$$
 \c T\=\{z=x+iy\in\cc4: y\in V'\}
=\c T_+\cup\c T_-, 
\e3 
where $\c T_\pm$ is the set of $z$'s with $y\in V_\pm'$.  The
fact that $V$ and $V'$ are almost identical is due to our arbitrary
choice $c=1$ for the speed  of light.  $V'$ is actually ``reciprocal'' to
$V$:  As $c$ increases, $V$ narrows and $V'$ widens.

A general function $\v F\0x$ with supp$\,\h\v F( p) \subset V$
therefore becomes ``polarized'' when extended to $\c T$: The
positive-frequency part $(p\in V_+)$ determines $\t\v F\0z$ in $\c
T_+$, and the negative-frequency part determines it in $\c T_-$.  
The positive and negative frequency components mix on the
common boundary  $\rr4$ of $\c T_+$ and $\c T_-$.  (The
boundaries of $\c T_\pm$ are seven-dimensional, but their
intersection is $\rr4$.)  If only real vector functions $\v F$ are
considered, then the positive and negative frequency parts are
``coupled'' by the reality condition $\ol{\h\v F(-p)}=\h\v F\0p$,
with the corresponding relation $\ol{\t\v F(\b z)}=\t\v F\0z$ in
$\c T$.   If $\v F$ is allowed to be complex-valued,  its positive-
and negative-frequency parts become independent.  In the case of
electrodynamics, we saw in the last section that they correspond to
the positive-helicity and negative-helicity parts of an
electromagnetic wave in the  $\v B+i\v E$ representation.  The
separation of helicities into $C_+$ and $C_-$ in Fourier space  is
translated, by the analytic-signal transform,  to their separation
into $\c T_+$ and $\c T_-$.

From a mathematical point of view, it suffices for $\h\v F\0p$ to
be supported in \st{any} double cone of the form $V_+\cup V_-$,
where $V_\pm$ are  convex cones intersecting only at their
common vertex.  Then $\t\v F$ is analytic at $x+iy$  whenever the
hyperplane orthogonal to $y$ separates $V_+$ and $V_-$, which
again means that $y$ belongs to the dual $V'$ of $V$, defined as
in \x(3.5).    The name ``analytic-signal transform'' derives from the
fact that when $n=m=1$ and $f$ is real-valued, then $\t f(x+iy)$
coincides with the ``analytic signal'' of $f$ for $y>0$, as first
defined by D. Gabor \ref[7] in his famous paper on communication 
theory.  In fact, if $n=1$, then $\t f(x+iy)$ is simultaneously the
analytic extension of the positive-frequency part of $f$ to the
complex upper half-plane and of the  negative-frequency part of
$f$ to the lower half-plane.   (These two half-planes now play the
roles of $\c T_+$ and $\c T_-$.)

 As already mentioned, $\t\v F\0x=\v F\0x$ formally, \ie by
setting $y=0$ in \x(3.2).  More precisely, $\v F$ is the
\st{boundary value} of $\v F$ in the sense that for any $y\in\rr
n\backslash\{0\}$,

$$
 \lim_{\ve\to 0^+}\[\t\v F(x+i\ve y)+\t\v F(x-i\ve y)\]
= 2\v F\0x \q \hbox{a.e.}
\e3 
 On the other hand, the ``jump'' of $\t\v F$ across
$\rr n$ is

$$ 
\lim_{\ve\to 0^+}\[\t\v F(x+i\ve y)-\t\v F(x-i\ve y)\]
= 2iH_y\,\v F\0x\q \hbox{a.e.,}
\e3 
 where  

$$ 
H_y\,\v F\0x
\=\rat1\ppp{\rm PV}\ir\rat{d\ttt}{\ttt}\ \v F(x-\ttt y) 
\e3 
 is the
\st{multidimensional Hilbert transform of $\v F$ in the direction of
$y\ne 0$} (Stein \ref[22]) and $PV$ denotes the Cauchy principal
value.

\sv4

\newcount\eq\eq=1 
\n{\bf 4. The  Electromagnetic Wavelets $\v\Y_z$}

\sv1

\n We are now ready to pursue our main theme, the construction
of the  electromagnetic wavelets and their \ru.  (For general
background on wavelet theory,  the reader may consult Chui
\ref[3], Daubechies \ref[5], Kaiser \ref[18] and the references
therein.)  Consider the extension of the electromagnetic field $\v
F\0x$ to the tube domain $\c T$ defined in \x(3.5) and \x(3.6).
Combining \x(2.21) and \x(3.2), we obtain

$$
 \t\v F(x+iy)
=\int_C\dpt\, 2\qqq( p\cdot y)\,e^{i p\cdot( x+iy)}\,
\v\P\0p\,\v f( p). 
\e4 
As earlier, assume that $\v f_\pm(\v p)\=\v f(\v p, \pm|\v p|)$ are
vector-valued Schwartz test functions,  to begin with.  Then $\t\v
F$ is analytic in $\c T$.   Fix an arbitrary $z=x+iy\in\c T$ (\ie
$y^2\=y_0^2-|\v y|^2>0$) and consider the linear operator $\c
E_z:\c H\to\cc3$ defined by $\c E_z\v F=\t\v F\0z$. This is an
\st{evaluation map} which, when applied to the field $\v F$, gives
the value of its extension at the complex space-time point $z$. 
Because of the analyticity of $\t\v F$,  $\c E_z$ turns out to be
bounded, as will be seen later. (It becomes unbounded as $y^2\to
0$.) We now define the electromagnetic wavelets as the adjoint
operators $\v\Y_z=\c E_z^*:\cc3\to\c H$.  To find these explicitly, 
choose any orthonormal basis $\v u_1, \v u_2, \v u_3$ of $\cc3$
and let $\v\Y_{z, k}\=\v\Y_z\v u_k\in\c H ,\, k=1, 2, 3$.  This
gives three solutions of Maxwell's equations,  all of which will be
wavelets ``at'' $z$.  $\v\Y_z$ is a \st{matrix-valued} solution of
maxwell's equations, obtained by putting the three (column) vector
solutions  $\v\Y_{z, k}$  together.  It will be convenient to
use the following ``star notation,'' intoduced in Kaiser [17]:  For any 
$\v F\in\c H$, let  $\v F^*:\c H\to\cx$ denote the linear functional
obtained by taking \ip s with $\v F$: 

$$
\v F^*\v G\=\la \v F, \v G\ra,\qq \v G\in\c H.
\e4
  $\v F^*$ is not to be confused
with the Hermitian transpose $\v F\0x^*$ of $\v F\0x\in\cc3$. 
Then the $k$-th component  of $\t\v F\0z$ \wrt{} the  basis $\{\v
u_k\}$ is

$$
\t  F_k\0z\=\v u_k^*\t\v F\0z=\v u_k^*\c E_z\v F
=\v u_k^*\v\Y_z^*\v F =(\v\Y_z\v u_k)^*\v F 
=\la \v\Y_{z, k}\,,\v F\ra. 
\e4
By \x(4.1),

$$
\v u_k^*\t\v F\0z
=\int_C\rat\dpt{p_0^2}\,2p_0^2\,\qqq( p\cdot y)\,e^{i p\cdot z}\, 
\v u_k^*\,\v\P\0p\,\v f( p)\,,
\e4
which shows that $\v\Y_{z, k}$ is given in the Fourier domain by

$$
\ybf_{z, k}\0p=
2p_0^2 \,\qqq( p\cdot y)\,e^{-i p\cdot \b z}\, \v\P\0p\,\v u_k\,.
\e4
Note that each  $\ybf_{z, k}\0p$ satisfies the constraint since
$\v\G\0p\,\v\P\0p=\v\P\0p$.   The matrix-valued wavelet
$\v\Y_z$  in the Fourier domain is therefore

$$
\ybf_z\0p=
2p_0^2 \,\qqq( p\cdot y)\,e^{-i p\cdot \b z}\, \v\P\0p .
\e4
In the space-time domain we  have (using 
$\v\P\0p\, \ybf_z\0p=\ybf_z\0p$)

$$
\v\Y_z(x')\=\int_C\dpt\,e^{ip\cdot x'}\, \ybf_z\0p
=\int_C\dpt \,2p_0^2 \,\qqq( p\cdot y)\,
e^{i p\cdot(x'-\b z)}\,\v\P\0p. 
\e4

Now that we have the wavelets, we want to make them into a
``basis'' that can be used to decompose and compose arbitrary
solutions.  This will be accomplished by constructing a ``resolution
of unity'' in terms of the wavelets.  To this end, we   derive an
expression for the \ip{} in $\c H$ directly in terms of the values
$\t\v F\0z$ of the extended fields.  To begin with, it will suffice to
consider the values of $\t\v F$ only at \st{Euclidean space-time
points,} \ie at points with a imaginary time coordinate
$z_0=is$ and real space coordinates $\v z=\v x$.    In order for $z$
to belong to $\c T$, it is only necessary to have $s\ne 0$.  We
denote the set of all such  points by $E$. The name
``Euclidean'' stems  from the fact that at such points, the negative
of the indefinite Lorentzian metric restricts to the positive-definite
Euclidean metric on $E$:  $-z^2=-(is)^2+|\v x|^2=s^2+|\v
x|^2$.   Later, $\v x$ will be interpreted as the center of the
wavelets $\v\Y_{z, k}\,$, and $s$ as their helicity and scale. 
Using \x(4.1) and letting $\ooo\=|\v p|=|p_0|$, we have

$$
\eqalign{ 
&\t\v F(\v x, is)
=\int_C \dpt
2\qqq(p_0s)\,e^{-p_0s-i\v p\cdot\v x} \,\v\P\0p\,\v f\0p\cr
&=2\int_\rr3\dpt\,e^{-i\sp px}\[ \qqq(\ooo s)
\,e^{-\ooo s}\,\v\P(\v p, \ooo)\,\v f(\v p, \ooo)
+ \qqq(-\ooo s)\,e^{\ooo s}\,\v\P(\v p, -\ooo)\,\v f(\v p, -\ooo)\]\cr 
&=\Bigl[\ooo\ii\qqq(\ooo s)\,e^{-\ooo s}\,
\v\P(\v p,\ooo)\,\v f(\v p,\ooo)
 + \ooo\ii\qqq(-\ooo s)\,e^{\ooo s}\, 
\v\P(\v p,-\ooo)\,\v f(\v p,-\ooo)\Bigr]^\vee(\v x), 
\cr}\e4 
where $^\vee$ denotes the inverse
\FT{} \wrt{} $\v p$.  Hence by Plancherel's formula, 

$$
\eqalign{ 
&\int_\rr3\d3x \,|\t\v F(\v x, is)|^2\cr
&=\int_\rr3\rat{\d3p}{\tp^3\ooo^2} 
\Bigl[\qqq(\ooo s)\,e^{-2\ooo s}\,
|\v\P(\v p,\ooo)\,\v f(\v p,\ooo)|^2
+ \qqq(-\ooo s)\,e^{2\ooo s} \,
|\v\P(\v p,-\ooo)\,\v f(\v p,-\ooo)|^2\Bigr],
\cr}\e4 
where we used
$\qqq\0u^2=\qqq\0u$ and $\qqq\0u\,\qqq(-u)=0$ for $u\ne 0$. 
Thus

$$
\eqalign{ 
\int_E\d3x\,ds \, |\t\v F(\v x, is)|^2
&=\int_\rr3\rat{\d3p}{2\tp^3\ooo^3}
 \Bigl[\, |\v\P(\v p,\ooo)\,\v f(\v p,\ooo)|^2
+ |\v\P(\v p,-\ooo)\,\v f(\v p,-\ooo)|^2\Bigr]\cr 
&=\int_C\rat\dpt{p_0^2}\,|\v\P\0p\,\v f\0p|^2
 =\int_C\rat\dpt{p_0^2}\,\v f\0p^*\,\v\P\0p\,\v f\0p
=\|\v F\|^2, 
\cr}\e4 
since $\v\P^*\v\P=\v\P^2=\v\P$.  Let $\t\c H$ be the set of all
analytic-signal transforms $\t\v F$ of solutions $\v F\in \c H$.
For $\t\v F, \t\v G\in\t\c H$, write

 $$
\la\t\v F, \t\v G\ra=\int_E\d3x\,ds\ \t\v F\0z^*\,\t\v G\0z.
\e4
Then \x(4.10)  leads immediately to the \ff{}   result.

\sv1

\proclaim Theorem 1.  $\t\c H$ is a Hilbert space under the \ip{}
\x(4.11), and the map $\v F\mt\t\v F$ is unitary from $\c H$ onto
$\t\c H$.

\sv2
\n\st{Proof:} By the  polarization identity, \x(4.10) implies 

$$
\la\t\v F, \t\v G\ra=\la\v F, \v G\ra,
\e4
so the map is an isometry. It is obviously surjective, by the
definition of $\t\c H$. \/\qed

\sv2 

\n With the ``star notation'' introduced earlier,   the  Hermitian
transpose $\t\v F\0z^*:\cc3\to\cx$ of the `column vector' $\t\v
F\0z\in\cc3$ is  the composition

$$
\t\v F\0z^*=(\v\Y_z^*\v F)^*=\v F^*\v\Y_z,
\e4
where  $\v F^*:\c H\to\cx$ denotes the linear functional \x(4.2). 
Hence the integrand in \x(4.11) is

$$
\t\v F(z)^*\,\t\v G(z) =(\v\Y_z^*\, \v F)^*\,\, \v\Y_z^*\, \v G 
=\v F^*\, \v\Y_z\,\v\Y_z^*\, \v G, 
\e4
where $\v\Y_z\,\v\Y_z^*: \c H\to\c H$ is the composition of
$\v\Y_z^*$ and $\v\Y_z\,$,  and \x(4.11) reads

$$ 
\int_E\d3x\,ds \  
\v F^*\, \v\Y_z\,\v\Y_z^*\, \v G 
=\v F^*\,\v G,\qq \v F, \v G\in\c H.
\e4

\sv2

\proclaim Theorem 2.  \hfill\break 
(a) The wavelets $\v\Y_z$ with $z\in E$
give the \ff{} resolution of the identity $I$ in $\c H$:
$$
\int_E\d3x\,ds \ \v\Y_z\,\v\Y_z^*=I,
\e4
where the equality holds in the weak topology of $\c
H$, \ie \x(4.15) is satisfied.
\hfill\break
(b)  Every solution $\v F\in \c H$ can be written as a superposition
of the wavelets $\v\Y_z$ with
$z=(\v x, is)\in E$, according to  
$$ 
\v F
=\int_E\d3x\,ds \ \v\Y_{\v x, is}\,\v\Y_{\v x, is}^*\,\v F
=\int_E\d3x\,ds \ \v\Y_{\v x, is}\,\t\v F(\v x, is), 
\e4
\ie
$$
\v F(x')=\int_E\d3x\,ds \ \v\Y_{\v x, is}(x')\,
\t\v F(\v x, is) \q\hbox{a.e.}
\e4
\x(4.17) holds  weakly in $\c H$ (\ie the \ip s of both sides
with any member of $\c H$ are equal).  However, for the {\rm
extended} fields, we have
$$
\t\v F(z')=\v\Y_{z'}^*\,\v F 
=\int_E\d3x\,ds \ \v\Y_{z'}^*\,\v\Y_{\v x, is}\,\t\v F(\v x, is) 
\e4
pointwise for all $z'\in\c T$.  

\sv2

\n\st{Proof:} Only the pointwise convergence in \x(4.19) remains to
be shown.  This follows from the boundedness of $\v\Y_{z'}$,
which will be proved in Section \x5.  \qed

\sv2

The pointwise equality fails, in general, for the boundary values
$\v F\0x$  because the evaluation maps (or, equivalently, their
adjoints $\v\Y_z$) become unbounded as $y\to 0$.  This will be
seen in the next section.

 The opposite composition $\v\Y_{z'}^*\v\Y_z: \cc3\to\cc3$ is a
matrix-valued function on $\c T\times\c T$:

$$\eqalign{  
\v K(z'\M \b z)
&\=\v\Y_{z'}^*\, \v\Y_z =\int_C\rat\dpt{p_0^2}\,
4p_0^4\,\qqq(p\cdot y')\,\qqq(p\cdot y)\, 
e^{ip\cdot (z'-\b z)}\, \v\P\0p^2\cr 
&=4\int_C\dpt\ p_0^2\,
\qqq(p\cdot y')\,\qqq(p\cdot y)\, 
e^{ip\cdot (z'-\b z)}\,\v\P\0p .
\cr}\e4
Eq.~\x(4.19) shows that $\v K(z'\M \b z)$ is  a \st{reproducing
kernel} for the \hs{} $\t\c H$;  see Kaiser \ref[11] for  background
and references.  The boundary value of $\v K(z'\M \b z)$ as $y'\to
0$ is, according to \x(3.7) and \x(4.7), given by

$$
\v K(x'\M \b z)=\rat12\lim_{\ve\to 0^+}\[\v K(x'+i\ve y'\M \b z)
+\v K(x'-i\ve y'\M \b z)\]=\v\Y_z(x').
\e4
Hence, to find the wavelets explicitly, we must compute their
reproducing kernel.  This is done in the next section.

The  meaning of the index $k$ in $\v\Y_{z, k}$ deserves to be
examined.  Since $\v\P\0p$ is the orthogonal projection  to
the eigenspace of $\v\G\0p$ with the nondegenerate eigenvalue 1, 
all the columns (as well as the rows) of $\v\P\0p$ are all multiples
of one another.  But  the coefficients are $p$-dependent,  and  the
algebraic linear dependence in Fourier space translates to a
\st{differential  equation} in space-time, relating the different
wavelets $\v\Y_{z, k}$.  For the columns, this differential equation
is just Maxwell's vector  equation \x(2.2). (Recall that the scalar
equation is then implied by the wave equation.)  Since $\v\P\0p$
is Hermitian, the same argument goes for the rows.  Explicitly,

$$
\v\G\0p\ybf_z\0p=\ybf_z\0p=\ybf_z\0p\v\G\0p.
\e4
When multiplied through by $p_0$ and transformed to
space-time, these read

$$
\nabla'\times\v\Y_z(x')=-i\pl_0'\v\Y_z(x')=
\v\Y_z(x')\times \overleftarrow{\nabla'},
\e4
where $\pl_0'$ denotes the partial \wrt{} $x_0'$, $\nabla'$ the 
gradient \wrt{} $\v x'$,  and $\overleftarrow{\nabla'}$
indicates that $\nabla'$ acts to the left, \ie on the column index. 
This states that \st{not only the columns, but also the rows of $\v
\Y_z$ are solutions of Maxwell's equations.  The three wavelets $\v
\Y_{z, k}$ are thus coupled.}  Note also that since $\v\Y_z(x')=\v
\Y_{z-x'}\00$, Eq.~\x(4.23) can  be rewritten as

$$
\nabla\times\v\Y_z=-i\pl_0\v\Y_z=
\v\Y_z\times \overleftarrow\nabla,
\e4
where $\pl_0$ and $\nabla$ are the corresponding operators
\wrt{} the \st{labels} $x_0=\re z_0$ and $\v x=\re \v z\,$.

We will see in Section \x7 that the reconstruction of $\v F(x')$ from
$\t\v F(\v x, is)$  can be obtained by a much simpler method than 
\x(4.18), using only a \st{ single scalar wavelet} $\Y_{\v x, is}(x')$
instead of the matrix wavelet $\v\Y_{\v x, is}(x')$  (or three
vector wavelets  $\v\Y_{\v x, is, k}(x')$). However, 
that presumes that we already know $\t\v F(\v x, is)$, and without
this knowledge the reconstruction becomes meaningless, since no
new solutions can be obtained this way.   \st{The use of 
matrix wavelets will be  necessary in order to give a
generalization of \x(4.17)--(4.19), where $\t\v F(\v x, is)$ can be
replaced with an unconstrained coefficient function.}   In other
words, we need matrix wavelets in space-time for exactly the
same reason that $\v\P\0p$ was needed in Fourier space 
(Eq.~\x(2.21)):   To eliminate the constraints in the coefficient
function.

\sv4

\newcount\eq\eq=1 
\n{\bf 5. The Reproducing Kernel}

\sv1

\n In order to obtain detailed information on the wavelets, we
 compute the reproducing kernel \x(4.20) explicitly.
Note, first of all, that if $y'\cdot y<0$ (\ie $z'\in \c T_+$ and $z\in \c
T_-$ or $z'\in \c T_-$ and $z\in \c T_+$),  then $\v K(z'\M \b z)=0$
since $p\cdot y'$ and $p\cdot y$ have opposite signs for all $p\in
C$.  Hence it suffices to compute the kernel for $z'$ and $z$ in the
same half of $\c T$.  Furthermore, $\v\P(-p)=\v\P\0p$ since
$\v\G(-p)=\v\G\0p$.  Hence, letting $z'\to\b z'$ and $z\to\b z$ in
\x(4.20)  gives

$$\eqalign{
 \v K(\b z'\M  z)
&=4\int_C\dpt\ p_0^2\,\qqq(-p\cdot y')\, \qqq(-p\cdot y)\, 
e^{ip\cdot (\b z'- z)}\,\v\P\0p\cr
&=4\int_C\dpt\ p_0^2\,\qqq(p\cdot y')\, 
\qqq(p\cdot y)\,
e^{ip\cdot ( z- \b z')}\,\v\P\0p 
=\v K(z\M  \b z'), 
\cr}\e5
 where the last equality is obtained by letting $p\to -p$.  Thus it
suffices to compute  the kernel for $z', z\in\c T_+$.  In this case,

$$ 
\v K(z'\M \b  z)
= 4\int_{C_+}\dpt\ p_0^2\, e^{ip\cdot ( z'- \b z)}\,\v\P\0p
 \=\v L( z'- \b z)
\e5
 is analytic in $w\= z'- \b z\in\c T_+$.   It can be shown
that $\v L(iy)$ with $y\in V_+'$ uniquely determines $\v L\0w$
for all $w\in\c T_+$ by analytic continuation, hence it suffices to
compute only $\v L(iy)$ for $y\in V_+'$.  Now the matrix elements
of $2p_0^2\,\v\P\0p$ are given by \x(2.20):

$$\eqalign{
2\,p_0^2\,\P_{mn}\0p
&=\ddd_{mn}p_0^2-p_mp_n
+i\sum_{k=1}^3\ve_{mnk}\,p_0p_k.
\cr}\e5
 To compute $\v L(iy)$,  it is useful to write the coordinates of
$y$ in \st{contravariant} form:  $y^0=y_0$, $y^m=-y_m\,
(m=1,2,3)$, so that $p\cdot y=\sum_{\mmm=0}^3
p_\mmm\,y^\mmm$.  Letting $\pl_\mmm$ denote the partial
derivative \wrt{} $ y^\mmm$ for $\mmm=0, 1, 2, 3$, we have

$$
 2\int_{C_+}\dpt\ p_\mmm\,p_\nnn\, e^{-p\cdot y}
=\pl_\mmm\,\pl_\nnn \,S(y)\=S_{\mmm\nnn}\0y,  
\qq\mmm, \nnn=0,1,2,3,
\e5
 where 

$$ 
S\0y\=2\int_{C_+}\dpt\ e^{-p\cdot y},\qq y\in V_+'\,.
\e5
 Thus \x(5.3) and \x(5.4) give the matrix elements of $\v L(i y)$ as

$$ 
L_{mn}(iy)=\ddd_{mn}S_{00}\0y-S_{mn}\0y
+i\sum_{k=1}^3\ve_{mnk}\,S_{0k}\0y,\q  m,n=1,2,3.
\e5
It only remains to compute $S\0y$. For this, we use the fact that 
$S\0y$ is invariant under Lorentz transformations, since $p\cdot
y$  and $\dpt$ are invariant and $C_+$ is a homogeneous space
for the proper Lorentz group.  Since $y\in V_+'$, there exists a
Lorentz transformation mapping $y$ to $(\v 0, \lll)$, where
$\lll\0y\=(y_0^2-|\v y|^2)^{1/2}>0$.  The invariance of $S$ implies
that $S(y)=S(\v 0, \lll)$.  Letting $\ooo=|\v p|$ again, we thus have

$$ 
S\0y=2\int_\rr3\rat{\d3p}{16\ppp^3 |\v p|}\,e^{-\lll|\v p|}
=\rat1{2\ppp^2}\int_0^\8\ooo \,d\ooo\,e^{-\ooo \lll}
=\rat1{2\ppp^2\lll^2}. 
\e5
Taking partials \wrt{} $y^\mmm$  and $y^\nnn$ gives

$$
 S_{\mmm\nnn}\0y 
=\rat{4y_\mmm y_\nnn-g_{\mmm\nnn} \lll^2}{\ppp^2\lll^6}
\,, \qq\mmm, \nnn=0,1,2,3, \e5
where $g_{\mmm\nnn}=$
diag$(1, -1, -1, -1)$ is the Lorentz metric. It \fs{} that

$$
L_{mn}(iy)=\rat2{\ppp^2\lll^6}
\Bigl[\ddd_{mn}(y_0^2+y_1^2+y_2^2+y_3^2)
-2y_my_n+2i\sum_{k=1}^3\ve_{mnk}y_0y_k\Bigr]. 
\e5
To compute $\v L\0w$ for $w\in\c T_+$, we need only replace $y$
with $-iw$.  This gives

$$ 
L_{mn}(w)=\rat2{\ppp^2w^6}
\Bigl[\ddd_{mn}(w_0^2+w_1^2+w_2^2+w_3^2)
-2w_mw_n+2i\sum_{k=1}^3\ve_{mnk}w_0w_k\Bigr],
\e5
where $w^6\= (w\cdot w)^3$. The full kernel is obtained by
setting $w=z'-\b z$ and multiplying by $\qqq(y'\cdot y)$, which
ensures that it vanishes when $z'$ and $z$ are in opposite
halves of $\c T$:

$$ 
\v K(z'\M \b z)=\qqq(y'\cdot y)\,\v L(z'-\b z),\qq z', z\in\c T.
\e5

In Section \x4 we stated that due to the analyticity of $\t\v F\0z$,
the evaluation maps $\c E_z$ (and with them, the wavelets $\v
\Y_z=\c E_z^*$) are bounded, and that they become unbounded as
$z=x+iy$ approaches the boundary of $\c T$, \ie $y^2\to 0$.  This
can now be verified by examining $\v K(z\M \b z)=\v \Y_z^*\,\v
\Y_z$.  By \x(5.11),   

$$
\v K(z\M \b z)=\qqq(y^2)\,\v L(2iy)=\v L(2iy)
\e5
for all $z\in\c T$, since $y^2>0$ in $V'$.  Eq.~\x(5.9)
shows that  $\v \Y_z^*\,\v \Y_z$ is indeed bounded when
 $z\in\c T$ diverges as $y^2\to 0$.  For example, if $y=(\v 0, s)$
(which can always be arranged by applying a Lorentz
transformation), then

$$
\v \Y_z^*\,\v \Y_z=\rat1{8\ppp^2\,s^4}\,I,
\e5
where $I$ is the identity matrix in $\cc3$.

\sv4

\n{\bf 6. Atomic Composition of Electromagnetic Waves}
\newcount\eq\eq=1   
\sv1

\n The reproducing kernel computed in the last section can be used
to \st{construct} electromagnetic waves according to local 
specifications, rather than merely to \st{reconstruct} known
solutions from their analytic-signal transforms on $E$.  This is
especially interesting because the Fourier method for constructing
solutions (Section \x2) uses plane waves and is therefore
completely unsuitable to deal with questions involving local
properties of the fields.  It will be shown in Section \x7 that the
wavelets $\v\Y_{x+iy}(x')$ are localized solutions of Maxwell's
equations,  at the ``initial'' time $x_0'=x_0$. Hence we call the
composition of waves from wavelets ``atomic.''

Suppose $\t\v F$ is the analytic-signal transform of a solution
$\v F\in \c H$ of Maxwell's equations.  Then according to \x(4.10),

$$
\int_E\d3x\,ds\ |\t\v F(\v x, is)|^2=\|\v F\|^2<\8. 
\e6
Let $\c L^2\0E$ be the set of \st{all} measurable functions $\v\F:
E\to\cc3$ for which the above integral converges.  $\c L^2\0E$ is a
\hs{} under the  obvious \ip, obtained from \x(6.1)
by polarization.  (In fact, we could identify $E$ with $\rr4$ and $\c
L^2\0E$ with $L^2(\rr4)$ since the set $\rr4\backslash E=\{(\v x,
0):\v x\in\rr3\}$ has zero measure in $\rr4$.  But this could
cause confusion between the Euclidean region $E$ and real
spacetime $\rr4$.)  Define the map $R_E:\c H\to\c
L^2\0E$ by 

$$
(R_E\v\, \v F)(\v x, is)\=\v\Y_{\v x, is}^*\v F=\t\v F(\v x, is).
\e6
That is, $R_E\,\v F$ is the restriction $\t\v F\,|_E$ to $E$  of the
analytic-signal transform $\t\v F$ of $\v F$. Then \x(6.1) implies
that the \st{range} $\c W$ of $R_E$ is a closed subspace of $\c
L^2\0E$, and $R_E$ maps $\c H$ isometrically onto $\c W$.   (In the
Physics literature, an operator which transforms fields in real
space-time to their counterparts in Euclidean space-time is called a
\st{Wick rotation.})  The following theorem characterizes the range
of $R_E$ and gives the adjoint $R_E^*$.

\vfill\eject

\proclaim Theorem \x3. \hfill\break
 (a) The range of $R_E$ is the set $\c W$ of all $\v\F\in\c L^2\0E$
satisfying the ``consistency condition''
$$
\v\F(z')=\int_E \d3x\, ds\,\v K(z'\M\v x, -is)\,\v \F(\v x, is),
\e6
pointwise in $z'\in E$. \hfill\break
(b) The adjoint operator $R_E^*: \c L^2\0E\to\c H$ is given by
$$
R_E^*\v \F=\int_E\d3x\, ds\,\v \Y_{\v x, is}\,\v\F(\v x, is),
\e6
where the integral converges weakly in $\c H$.

\sv2

\n\st{Proof:}  If $\v \F\in\c W$, then $\v\F(\v x, is)=\t\v F(\v x,
is)$ for some $\v F\in\c H$, and \x(6.3) reduces to \x(4.19), which
holds pointwise in $z'\in E$.   On the other hand, given a function
$\v\F\in\c L^2\0E$ which satisfies \x(6.3), let $\v F$ denote the
\rhs{} of \x(6.4).  Then  for any $\v G\in\c H$,

$$
\v G^*\v F =\int_E\d3x\, ds\,\t\v G(\v x, is)^* \v\F(\v x, is)
=\la R_E\,\v G, \v\F\ra_{\c L^2}\,,
\e6
where we have used $\v G^*\v\Y_{\v x, is}=
(\v\Y_{\v x, is}^*\v G)^*=\t\v G(\v x, is)^*$.  Hence the integral in
\x(6.4) converges weakly in $\c H$. The transform of $\v F$ under
$R_E$ is

$$\eqalign{
(R_E\,\v F)(z')&=\v\Y_{z'}^*\v F=
\int_E\d3x\, ds\,\v\Y_{z'}^*\v\Y_{\v x, is} \v\F(\v x, is)\cr
&=\int_E\d3x\, ds\,\v K(z'\M \v x, -is)\, \v\F(\v x, is)=\v\F (z'),
\cr}\e6
by \x(6.3). Hence $\v\F\in\c W$ as claimed, proving (a). 
Eq.~\x(6.5)  states that
$\la\v G,\v F\ra_\c H=\la R_E\,\v G, \v\F\ra_{\c L^2}$. That shows
that $\v F=R_E^*\v\F$, proving (b).  \sh2\qed

\sv2

Eq.~\x(6.4) constructs a solution $R_E^*\v\F\in\c H$ from a
coefficient function $\v \F\in\c L^2\0E$.  When $\v\F$ is actually
the transform $R_E\,\v F$ of a solution $\v F\in\c H$, then 
$\v\F(\v x, is)=\v\Y_{\v x, is}^*\v F$ and

$$
R_E^*\v \F
=\int_E\d3x\, ds\,\v \Y_{\v x, is}\,\v\Y_{\v x, is}^*\v F=\v F,
\e6
by \x(4.17).  Thus $R_E^*R_E=I$, the identity in $\c H$.  (This is
equivalent to \x(6.1).)  We now examine the opposite composition.

\sv2

\proclaim Theorem \x4.  The orthogonal projection to $\c W$ in 
$\c L^2\0E$ is  the composition \hfill\break
 $P\=R_ER_E^*: \c L^2\0E\to\c L^2\0E$, which is given by
$$
(P\v\F)(z')\=\int_E\d3x\, ds\,\v K(z'\M \v x, -is)\,\v\F(\v x, is).
\e6
\sv2

\n\st{Proof:}  By \x(6.4), 

$$\eqalign{
(R_ER_E^*\v\F)(z')&\=\v\Y_{z'}^*R_E^*\v\F
=\int_E\d3x\,ds\,\v\Y_{z'}^* \v\Y_{\v x, is}\v\F(\v x, is)
=(P\v\F)(z'),
\cr}\e6
since $\v\Y_{z'}^* \v\Y_{\v x, is}=\v K(z', \v x, -is)$.  Hence
$R_ER_E^*=P$.  This also shows that $P^*=P$.   Furthermore, 
$R_E^*R_E=I$ implies that $P^2=P$, hence $P$ is indeed  the
orthogonal projection to its range.  It  only remains to show that
the range of $P$ is $\c W$.   If $\v \F=R_E\,\v F\in\c W$, then
$R_ER_E^*\v\F=R_ER_E^*R_E\,\v F=R_E\,\v F=\v\F$.      Conversely, 
any function in the range  of $P$ has the form
$\v\F=R_ER_E^*\v\Q$ for some $\v\Q\in\c L^2\0E$, hence
$\v\F=R_E\v F$ where $\v F=R_E^*\v\Q\in\c H$.\sh2\qed

\sv2

When the coefficient function $\v\F$ in \x(6.4) is the transform of
an actual solution, then $R_E^*$ \st{reconstructs} that solution. 
However, this process does not appear to be too interesting, since 
we must have a complete knowledge of $\v F$ to compute $\t\v
F(\v x, is)$.   For example, to compute $\t\v F(\v x, is)$ by \x(3.1),
we must know $\v F(\v x, t)$ for all $\v x$ and all $t$.  Hence, no
``initial-value problem'' is solved by \x(6.4) when applied to
$\v\F\in\c W$.     However, the option of applying \x(6.4) to
\st{arbitrary} $\v\F\in\c L^2\0E$ is a very attractive one, since it
is guaranteed to produce a solution without any assumptions on
$\v\F$ other than square-integrability. It is appropriate to call
$R_E^*$ the \st{construction operator} associated with the \ru{}
\x(4.16).  It  can be used to construct solutions in $\c H$ from
\st{unconstrained} functions  $\v\F\in\c L^2\0E$.  In fact, it is
interesting to compare the wavelet construction formula

$$
\v F(x')=\int_E\d3x\,ds\,\v\Y_{\v x, is}(x')\,\v\F(\v x, is)
\e6
directly with its Fourier counterpart \x(2.21):

$$
\v F(x')=\int_C\dpt e^{ip\cdot x'}\,\v\P\0p\,\v f\0p.
\e6
In both cases, the coefficient functions  ($\v\F$ and $\v f$) are
unconstrained (except for the respective square-integrability
requirements).   The building blocks in \x(6.10) are the
matrix-valued wavelets parameterized by $E$, whereas those in
\x(6.11) are the matrix-valued plane-wave solutions $e^{ip\cdot
x'}\,\v\P\0p$ parameterized by $C$.  

\sv4

\n{\bf 7. Interpretation of the Wavelet  Parameters}
\newcount\eq\eq=1   
\sv1

\n Our goal in this section is twofold:  (a)  Reduce the wavelets
$\v\Y_z$ to a sufficiently simple form that they can actually be
visualized, and (b) use the ensuing picture to give a complete
physical and geometric interpretation of the eight complex
space-time parameters $z\in\c T$ labeling $\v\Y_z$.  That the
wavelets can be visualized at all is quite remarkable, since
$\v\Y_z(x')$ is  a complex matrix-valued function of $x'\in\rr4$
and $z\in\c T$.   However, the symmetries of Maxwell's equations 
can be used to reduce the number of effective variables one by one,
until all that remains is a single complex-valued function of two
real variables, whose real and imaginary parts
can be graphed separately.

We begin by showing that the parameters $z\in\c T$ can be
eliminated entirely.   Recall that  $\v\Y_z(x')$ is the boundary
value of the reproducing  kernel, according to \x(5.11) and \x(3.7):

$$\eqalign{
\v\Y_z(x')&=\rat12\,\lim_{\ve\to 0}
\[\v K(x'+i\ve y'\M\b z)+\v K(x'-i\ve y'\M\b z)\]\cr
&=\rat12\,\lim_{\ve\to 0}
\[\qqq(y'\cdot y)\v L(x'+i\ve y'-\b z)
+\qqq(-y'\cdot y)\v L(x'-i\ve y'-\b z)\]\cr
&=\rat12\,\[\qqq(y'\cdot y)
+\qqq(-y'\cdot y)\]\v L(x'-\b z)
=\rat12\,\v L(x'-\b z).
\cr}\e7
Hence

$$
\v \Y_{x+iy}(x')=\rat12\,\v L(x'-x+iy)=\v \Y_{iy}(x'-x),
\e7
and  $\v\Y_{x+iy}$ is a translated version of $\v\Y_{iy}\,$.  It
therefore suffices to examine only $\v\Y_{iy}$ with $y\in V'$.
Eq.~\x(4.7), combined with $\v\P(-p)=\v\P\0p$,  shows that 
$\v\Y_z(x')^*=\v\Y_{\b z}(x')$, hence
 it suffices to look only at $y\in V_+'$.   To reduce the number of
parameters still further, we use the fact that Maxwell's equations
are invariant under Lorentz transformations, and this invariance
implies certain transformation properties for the wavelets.   The
covariance  of the wavelets under the Lorentz group and, more
generally, under the  conformal group, will be studied in detail
elsewhere.  Here we remark only that Lorentz transformations
relate all the wavelets with equal values of $y^2$, hence it suffices
to study only $\v\Y_{iy}$ with
 $y=(\v 0, s)$ and $s>0$.  The physical significance of this will be
discussed below.  Finally,  note that $\v\G(ap)=\v\G\0p$ for any
$a>0$, since $\v v(ap)\=a\v p/ap_0=\v v\0p$.    Hence
$\v\P(ap)=\v\P\0p$, and \x(4.7) implies that 

$$
\v\Y_{\v 0, is}(x')=s^{-4}\,\v\Y_{\v 0, i}(x'/s).
\e7
Thus \st{all the wavelets $\v\Y_z,\, z\in\c T$,  can be obtained
by space-time translations, Lorentz transformations and scalings 
 from the single ``mother wavelet''}

$$\eqalign{
\v\Y\0x\=\v \Y_{\v 0, i}\0x
=2\int_{C_+}\dpt\,p_0^2\,e^{-p_0}\,e^{ip\cdot x}\,\v\P\0p.
\cr}\e7
 (Of course, \st{any} one of the $\v \Y_z$'s can
equally be chosen as the mother!)     In particular, the
wavelets parameterized by $(\v x, is)\in E$ are 

$$
\v\Y_{\v x, is}(\v x', t')
=s^{-4}\,\v\Y\(\rat{\v x'-\v x}s\,, \rat{t'}s\). 
\e7
Let $[\v\Y(\v x, t)]_{mn}$ denote the matrix elements of 
$\v\Y(\v x, t)$.  By \x(5.10), with $w_0=t+i,\ \v w=\v x$ and 
$r=|\v x|$, we have

$$
[\v\Y(\v x, t)]_{mn}=\rat1{\ppp^2}\,
\rat{\ddd_{mn}[(t+i)^2+r^2]-2x_mx_n
+2i(t+i)\sum_{k=1}^3\ve_{mnk}x_k}
{[(t+i)^2-r^2]^3}\,.
\e7
This is still a complex matrix-valued function in $\rr4$,
hence impossible to visualize directly.  We now eliminate the
polarization degrees of freedom.  Returning to the Fourier
representation of solutions, note that if  $\v f\0p$ already
satisfies the constraint \x(2.12), then $\v\P\0p\,\v f\0p=\v f\0p$
and \x(4.10) reduces to

$$
\int_E\d3x\,ds \, |\t\v F(\v x, is)|^2
=\int_C\rat\dpt{p_0^2}\,|\v f\0p|^2 =\|\v F\|^2 .
\e7 
Define the \st{scalar} wavelets by

$$
\Y_z(x')\=\int_C\dpt\,2p_0^2\qqq(p\cdot y)\,
e^{-p\cdot(x'-\b z)}
\e7
and the corresponding scalar kernel $K:\c T\times\c T\to\cx$ by

$$
K(z'\M \b z)=\int_C\dpt\, 4p_0^2\qqq(p\cdot y')\,\qqq(p\cdot y)\,
e^{ip\cdot(z'-\b z)}.
\e7
Then \x(7.7), with essentially the same argument as in the proof
of Theorem \x1, now gives the relations

$$\eqalign{
\t\v F(z')&=\int_E\d3x\,ds \  K(z'\M\v x, -is)\,
\t\v F(\v x, is) \q\hbox{pointwise in}\ z'\in\c T, \cr
\v F(x')&=\int_E\d3x\,ds \  \Y_{\v x, is}(x')\,
\t\v F(\v x, is) \q\hbox{a.e.} \cr
\v F&=\int_E\d3x\,ds \ \Y_{\v x, is}\,\t\v F(\v x, is)
\q\hbox{weakly in}\ \c H.
\cr}\e7
The first equation states that  $K(z'\M \b z)$ is still
a reproducing kernel on the range $\c W$ of $R_E: \c H\to\c
L^2\0E$.  The second and third equations state that an arbitrary
solution  $\v F\in\c H$ can be represented as a superposition of
the scalar wavelets, with $\t\v F=R_E\,\v F$ as a (vector)
coefficient function.  Thus, when dealing with coefficient
functions in the range $\c W$ of $R_E$, it is unnecessary to use
 the matrix-valued wavelets.  The main advantage of the latter
(and a very important one) is that they can be used even when the
ceofficient function $\v\F(\v x, is)$ does \st{not} belong to the
range $\c W$ of $R_E$, since they \st{project} $\v\F$ to $\c W$.

 The scalar  wavelets and kernel were introduced and studied in
Kaiser [12, 13].  They cannot, of course, be solutions of Maxwell's
equations precisely because they are scalars.  But they do satisfy
the wave equation, since every component of $\v\Y_z$ does so. 
To see their relation to the corresponding matrix quantities, note
that $\v\P\0p$ is a projection operator of rank 1, hence Trace
$\v\P\0p=1$.  Taking the trace on both sides of
 Eqs.~\x(4.7) and \x(4.20) therefore gives

$$
\Y_z(x')={\rm Trace\ }\v\Y_z(x'),\qq
K(z'\M \b z)={\rm Trace\ }\v K(z'\M\b z).
\e7
Taking the trace amounts, roughly,  to averaging over
polarizations. The trace of the mother wavelet $\v\Y$ is

$$
{\rm Trace\ }\v\Y\0x=\rat1{\ppp^2}\,
\rat{3(t+i)^2+r^2}{[(t+i)^2-r^2]^3}\=\Y(r, t),\q r\=|\v x|.
\e7
Because it is spherically symmetric, $\Y(r, t)$ can be easily
plotted.   Its real and imaginary parts are shown in Figs.~1 and
2.  These figures confirm that $\Y$ is a spherical wave  converging
towards the origin as $-\8<t<0$, becoming localized in a sphere of
radius $\sqrt{3}$ around the origin at $t=0$, and then diverging
away from the origin as $0<t<\8$.  Figure 3 shows $\Y(r, 0)$,
which is real.  Even though $\Y(r, 0)$ does not have
compact support (it decays as $r^{-4}$), it is seen to be very well
localized in $|\v x|\le\sqrt{3}$.

Now that we have a reasonabe interpretation of 
$\v\Y_{x+iy}$ with $y=(\v 0, s)$, let us return to interpret the
wavelets with $\v y\ne 0$.  Suppose $y=(\v y, s)\in V_+'\,$, and 
let  $\v v\=\v y/s$.  Then $y\in V_+'$ implies  $|\v v|<1$. (We have
chosen units of length and time in which the speed of light $c=1$; 
for general units, $|\v v|<c$.)  Hence we can perform a Lorentz
transformation to a reference frame moving with velocity $-\v v$
relative to the original frame.  In the new frame, $\v y$ has
coordinates $y'=(\v 0, \sqrt{s^2-|\v y|^2})$, hence our wavelet has a
stationary center.  Returning to the original frame, we conclude
that $\v\Y_{x+iy}$ is a wave whose center is moving with the
uniform velocity $\v v=\v y/s$.  It is a \st{Doppler-shifted}
version of the stationary wavelet with $\v y'=\v 0$.  Thus  \st{each
of the eight real parameters  $z=(\v x, t)+i(\v y, s)\in\c T$ has a 
physical and geometric significance:  $\v x$ and $t$ give the
location and time at which $\v\Y_z$ is localized;  $\v v=\v y/s$
gives the velocity of its center; $|s|$ gives its scale (width at time
$t$), and the sign of $s$ gives its helicity.}   Since all these
parameters, as well as the wavelets which they label, were a direct
consequence of the extension of the electromagnetic field to
complex space-time,  it would appear that $\c T$  is a  natural
arena in which to study electrodynamics.

\vfill\eject

\n{\bf 8. Moving and Accelerating Wavelet Representations}
\newcount\eq\eq=1

\sv1

\n  The construction of the electromagnetic wavelets has
been completely unique, in the following sense:  (a)  The \ip{}
\x(2.31) on  solutions is uniquely determined, up to a constant
factor, by the requirement that it be Lorentz-invariant. 
 (b)  The analytic extensions of the positive- and
negative-frequency parts of $\v F$ to $\c T_+$ and $\c T_-\,$,
\resp, are certainly unique, hence so is $\t\v F\0z$.  (c) The
evaluation maps $\c E_z\v F=\t\v F\0z$ are unique, hence so are
their adjoint $\v\Y_z\=\c E_z^*$.    On the other hand, the choice of
the Euclidean space-time region $E$ as a parameter space for
expanding solutions is rather arbitrary.  $E$ may be regarded as
the group of space translations and scalings,  acting on real
space-time by $g_{\v x, is}x'=sx'+(\v x, 0)$. As such, it is a
subgroup of the conformal group $\c C$, which consists of
space-time translations, scalings, space rotations, Lorentz
transformations and special conformal transformations.
 $E$ is invariant under space rotations but not  under time
translations, Lorentz transformations or special conformal
transformations.  This non-invariance  can be exploited by
applying any of the latter  transformations to the \ru{}
\x(4.16) and using the transformation properties of the wavlelets. 
The general idea is that when $g\in \c C$ is applied to \x(4.16),
then another such \ru{} is obtained in which $E$ is replaced by its
image $gE$ under $g$.  If $gE=E$,  nothing new results.
If $g$ is a time translation, then the wavelets parameterized by
$gE$ are all localized at some time $t\ne 0$ rather than $t=0$. If
$g$ is a Lorentz transformation, then all wavelets with $z\in gE$
have centers which move with a uniform non-zero velocity rather
than being stationary.  Finally, if $g$ is a special conformal
transformation, then $gE$ is a curved submanifold of $\c T$ and
the wavelets parameterized by $gE$ have centers with varying
velocities.  This is consistent with results obtained by
 Page \ref[20] and Hill \ref[9], who showed that special conformal
transformations  can be interpreted as mapping to an
\st{accelerating} reference frame.

As a possible application,   consider an electromagnetic pulse
reflected  or emitted by a moving object.  After the reflection time,
and far away from boundaries, the pulse may be approximated by
a solution of  Maxwell's equations in free space, hence it can be
analyzed as in Section \x4 using  wavelets with stationary centers. 
However,  the analysis is likely to be more efficient (\ie have
fewer significant coefficients) if it is made in the reference frame
in which the reflecting object is at rest, with the reflection time as
the initial time of localization.  From the viewpoint of the receiver,
this means that  a representation with ``co-moving''  wavelets
should be used instead of one with stationary centers.  The details
will be presented elsewhere.

\sv4

I thank R.~F.~Streater for his hospitality at King's College, where
we had some helpful discussions concerning the helicity of the
electromagnetic wavelets.

\vfill\eject

\refs

[1] V.~Bargmann and E.~P.~Wigner, Group-theoretical discussion
of relativistic wave equations,  Proc.~Natl.~Acad.~Sci.~U.~S.
{\bf 34} (1948) 211-233.

[2]  H.~Bateman, The transformation of the electrodynamical
equations,  Proc.~London Math.~Soc.~{\bf 8} (1910) 223-264.

[3] C.~K.~Chui, \st{An Introduction to Wavelets,} Academic Press,
1992.

[4]  E.~Cunningham, The principle of relativity in
electrodynamics and an extension thereof,  Proc.~London Math.~Soc.
{\bf 8} (1910) 77-98.

[5]  I.~Daubechies, \st{Ten Lectures on Wavelets,} SIAM,
1992.

[6] A.~Einstein, H.~A.~Lorentz,  H.~Weyl and H.~Minkowski,  \st{The
Principle of Relativity,}  Dover, 1923.

[7]  D.~Gabor, Theory of communication, Proc.~IEE (London),
Ser.~3, {\bf 93} (1946)  429-457.

[8]  L.~Gross,  Norm invariance of mass-zero equations
under the conformal group, \break
 J.~Math.~Phys.~{\bf 5} (1964) 687-695.

[9]  E.~L.~Hill, On accelerated coordinate systems in
classical and relativistic mechanics, Phys.~Rev.~{\bf 67} (1945)
358-363;  On the kinematics of uniformly accelerated motions and
classical electromagnetic theory, {\sl ibid.\/} {\bf 72} (1947)
143-149; The definition of moving coordinate systems in
relativistic theories, {\sl ibid.\/}  {\bf 84} (1951) 1165-1168.

[10] J.~D.~Jackson, \st{Classical Electrodynamics,} Wiley, 1975.

[11]  G.~Kaiser, \st{Quantum Physics, Relativity, and Complex
spacetime: Towards a New Synthesis,} North-Holland,
Amsterdam, 1990.

[12]  G.~Kaiser, Wavelet electrodynamics, 
Physics Letters A {\bf 168} (1992) 28-34.

[13]  G.~Kaiser,  in \st{Progress in Wavelet Analysis and
Applications,}  Y.~Meyer and \break
S.~Roques, eds., Editions
Frontieres, 1993.

[14]  G.~Kaiser, Space-time-scale analysis of
electromagnetic waves, in Proc.~of IEEE-SP Internat.~Symp.~on
Time-Frequency and Time-Scale Analysis, Victoria  1992.

[15]  G.~Kaiser and R.~F.~Streater, Windowed Radon
transforms, analytic signals and the wave equation, in {\sl
Wavelets---A Tutorial in Theory and Applications,\/}  C.~K.~Chui,
ed., Academic Press, New York, 1992.

[16]  G.~Kaiser,  Phase-space approach to relativistic
quantum  mechanics.  \hfil\break
Part I: Coherent-state representation  of
the Poincar\'e group,  J.~Math.~Phys.~{\bf 18} (1977) 952-959; part
II: Geometrical Aspects, {\sl ibid.\/}  {\bf 19} (1978) 502-507.

[17]  G.~Kaiser,  Quantized fields in complex space-time, Ann.
Phys.~{\bf 173} (1987) 338-354.

[18] G.~Kaiser, \st{A Friendly Guide to Wavelets,} Birkh\"auser,
Boston, 1994.

[19] H.~E.~Moses, \st{Eigenfunctions of the curl operator,
rotationally invariant Helmholtz theorem, and applications to
electromagnetic theory and fluid mechanics,}  
SIAM J.~Appl.~Math.{\bf 21} (1971)  114-144.

[20]  L.~Page, A new relativity, Phys.~Rev.~{\bf 49} (1936)
254-268.

[21]  W. R\" uhl, Distributions on Minkowski space and their
connection with analytic representations of the conformal group,
Commun.~ math.~Phys.~{\bf 27} (1972) 53-86.

[22]  E.~Stein, \st{Singular Integrals and Differentiability
Properties of Functions,} Princeton University Press, 1970.

[23] E.~Stein and G.~Weiss, \st{Fourier Analysis on Euclidean
Spaces,} Princeton University Press, 1971.

[24] T.~Takiguchi, ``The windowed Radon transform for
distributions,''  Univ. of Tokyo preprint, 1994.

\bye